\documentclass{aa}
\usepackage{graphicx}
\def\H2{H$_2$}
\def\fH2{f_{\rm H_2}}
\def\nh{N_{\rm H}}
\def\nmol{N_{\rm H_2}}
\def\rd{{\rm d}}
\begin{document}
\title{Search for high column density systems with
gamma ray bursts}
\author{H. Hirashita \inst{1,2}
\fnmsep\thanks{Postdoctoral Fellow of the Japan Society for the
Promotion of Science (JSPS). Present address:
Center for Computational Sciences, University of Tsukuba,
                Tsukuba 305-8577, Japan},
H. Shibai \inst{2},
         \and
T. T. Takeuchi \inst{3}\fnmsep\thanks{Postdoctoral Fellow of
the JSPS for Research Abroad.
Present address: Astronomical Institute, Tohoku University,
Aramaki, Aoba-ku, Sendai 980-8578, Japan}
}
\offprints{H. Hirashita,\\ \email{hirashita@u.phys.nagoya-u.ac.jp}}
\institute{SISSA/International School of Advanced Studies,
    Via Beirut 4, 34014, Trieste, Italy
\and
Graduate School of Science, Nagoya University, Furo-cho,
                Chikusa-ku, Nagoya, 464-8602, Japan\\
\email{hirashita, shibai@u.phys.nagoya-u.ac.jp}
        \and
Laboratoire d'Astrophysique de Marseille, Traverse du Siphon BP
    8, 13376 Marseille Cedex 12, France\\
\email{tsutomu.takeuchi@oamp.fr}
}
\date{24 February 2006}
\abstract{
We investigate the possibility to search for metal-poor
high column density ($\ga 10^{23}~{\rm cm}^{-2}$) clouds at
high redshift ($z$) by using gamma ray burst (GRB)
afterglows. Such clouds could be related to primeval
galaxies which may cause a burst of star formation.
We show that a large part of hydrogen is in molecular form
in such a high column density environment. Therefore,
hydrogen molecules (\H2) rather than hydrogen atoms should
be searched for. Then we show that infrared \H2 lines are
detectable for metal-poor ($\la 0.01$ solar metallicity)
high column density ($\log\nh~[{\rm cm}^{-2}]\ga 23.5$)
systems at high-$z$ without suffering dust extinction.
Optical properties of dust in infrared could also be
constrained by observations of high column density systems.
Some possible scenarios
of producing high column density systems are finally
discussed in the context of galaxy evolution.
 }
\titlerunning{Searching for H$_2$ absorption with GRB afterglows}
\authorrunning{H. Hirashita et al.}
\maketitle
%


\section{Introduction}\label{sec:intro}

The intergalactic and interstellar gas is a main
component of the baryonic matter, and occupies most
of the volume in the Universe. Gas with various
column densities has been detected as line
absorptions of the background Quasi-Stellar
Objects (QSOs). The strong
Ly$\alpha$ absorption feature is often
used to identify absorbing clouds, which are
called Ly$\alpha$ clouds. In particular,
if the H\,{\sc i} column density is larger
than $2\times 10^{20}~{\rm cm}^{-2}$, the
absorbing system is called damped Ly$\alpha$
clouds (DLAs) because the continuum of background
QSOs is completely damped by the Ly$\alpha$
absorption
(e.g., Wolfe et al.\ \cite{wolfe86}).

So far the hydrogen column density has been
sampled up to $\log\nh\la 22$ (in this paper, we
show all the
column densities in units of cm$^{-2}$). It is
difficult to detect systems with higher column
density, since dust severely extinguishes the
background objects. Fall et al.\ (\cite{fall89})
show that QSOs behind DLAs tend to be redder than
those without DLAs.
This result strongly suggests that the dust
contained in high
column density systems extinguishes the
background light (but see
Murphy \& Liske \cite{murphy04}). If
$\log\nh\ga 22$, the effects of extinction
become significant and a large fraction of
absorption systems can be missed
(e.g., Vladilo \& P\'{e}roux \cite{vladilo05}).

Schaye (\cite{schaye01}) proposes another
possibility of missing high column density
systems. He shows that a large fraction of hydrogen
is in molecular form if $\log\nh\ga 22$. Such
molecular clouds may not be sampled by
Ly$\alpha$ lines of H\,{\sc i}, and thus, it is
necessary to search for molecular hydrogen (\H2).
Shibai et al.\ (\cite{shibai01}, hereafter S01)
investigate the observability of \H2 in absorption
for high column density systems. They focus on
near-infrared (NIR) vib-rotational lines, because
dust extinction is less severe in the NIR than
in the optical and ultraviolet (UV). Considering
typical QSOs as background sources, they show that
the high column density systems with
an \H2 column density of
$\log\nmol\ga 23$ could be
observed without being extinguished by dust if
the dust-to-gas ratio is less than 1 percent of
the Galactic value.

However, QSOs are usually sampled in the optical
wavelength, where dust extinction is severe.
Ellison et al.\ (\cite{ellison01}) investigate a
sample of radio-selected QSOs and conclude that
dust-induced bias in optical samples is small.
Therefore, the significance of
extinction bias is still debated.
In this paper, we consider the possibility to
use gamma-ray burst (GRB) afterglows as background
sources. Since GRB samples are defined in
wavelength not affected by dust extinction,
we expect less extinction bias in GRB samples than
in QSO samples. Indeed some DLAs are found to
lie in the sightline of GRBs (e.g.,
Vreeswijk et al.\ \cite{vreeswijk04}) and
some GRBs are surrounded by high column density
gas (e.g., Piro et al.\ \cite{piro05}).
Therefore,
it may be possible to use NIR continua of GRB
afterglows to detect NIR \H2 lines.

Recently, Inoue et al.\ (\cite{inoue06}) have
investigated the possibility of detecting the
absorption lines originating from protostellar cores
by using
background GRB afterglows. They find that radio
observations of some metal absorption lines of GRB
afterglows are useful to detect high-$z$ protostellar
gas. However, it is still interesting to directly
detect \H2 because \H2 is considered to play a
central role in star formation in metal-poor
environments (e.g.,
Mizusawa et al.\ \cite{mizusawa04}).
Therefore, in spite of its difficulty,
the possibility to detect IR \H2 lines is
worth investigating.

Since \H2 vib-rotational lines are generally weak,
large infrared (IR) telescopes are necessary. The
strongest vib-rotational \H2 lines are normally those
around the wavelength of
$\lambda\sim 2~\mu$m. Those lines shift to the mid-IR
(MIR) regime if we observe high-$z$ clouds. In MIR,
it is essential to avoid the emission of atmosphere
and telescope. Thus, a cooled space telescope is
ideal. There are indeed suitable future missions planned
currently such as the
{\it Space Infrared Telescope for Cosmology and Astrophysics}
({\it SPICA})\footnote{http://www.ir.isas.ac.jp/SPICA/index.html}.
{\it Herschel}\,%
\footnote{http://astro.estec.esa.nl/SA-general/Projects/Herschel/}
is also a planned space mission in IR (but not cooled),
and the same kind of targets can be accessible if
the background sources are luminous enough.
In NIR ($\lambda\la 5~\mu$m), the
{\it James Webb Space Telescope}
({\it JWST}\,)\footnote{http://ngst.gsfc.nasa.gov/}
is more sensitive than the {\it SPICA}, and
is appropriate for \H2 absorption at $z\la 1.5$.

First, in Sect.\ \ref{sec:model} we explain the method
to calculate \H2 absorption line luminosity.
Our results are shown in Sect.\ \ref{sec:result}, and
some relevant physical and observational
issues are discussed in Sect.\ \ref{sec:strategy}.
Our conclusions are given in
Sect.\ \ref{sec:conclusion}.

\section{H$_2$ absorption lines}\label{sec:model}

\subsection{Absorption line flux}
\label{subsec:abs_flux}

We follow S01 for the formulation concerning IR \H2
lines. We assume a uniform, cool gas cloud satisfying
$k_{\rm B}T_{\rm ex}\ll h\nu$, where $k_{\rm B}$
is the Boltzmann constant, $T_{\rm ex}$ is the
excitation temperature, $h$ is the
Planck constant, and $\nu$ is the frequency of an
\H2 line.

The optical depth of the line absorption,
$\tau_{\rm line}$, is estimated as
\begin{eqnarray}
\tau_{\rm line}\simeq\frac{\lambda^3}{8\pi}
\left(\frac{g_u}{g_\ell}\right) A_{u\ell}N_\ell
\frac{1}{\Delta v}\, ,\label{eq:tau_line}
\end{eqnarray}
where subscripts $u$ and $\ell$ indicate the
upper and lower levels of a transition, $g_u$
and $g_\ell$ are the degeneracy of each state,
respectively, $A_{u\ell}$ is Einstein's $A$
coefficient, $N_\ell$ is the column density of
the molecules in the lower state, and $\Delta v$
is the line width in units of velocity.
If the instrumental wavelength resolution is
larger than the physical line width, we should
take $\Delta v=c/R$, where $R$ is the resolution
defined by
$R\equiv\lambda /\Delta_{\rm i}\lambda$
($\Delta_{\rm i}\lambda$ is the wavelength
resolved by the instrument).

Next, we consider extinction by dust grains,
whose optical depth at the wavelength $\lambda$
is denoted as $\tau_{\rm dust}(\lambda )$. This
is estimated as
(Hirashita et al.\ \cite{hirashita05})
\begin{eqnarray}
\tau_{\rm dust}(\lambda )=\mu m_{\rm H}\nh
{\cal D}\langle\sigma_{\rm d}(\lambda )/m_{\rm d}
\rangle\, ,\label{eq:tau_dust}
\end{eqnarray}
where $\mu$ is the mean atomic mass per H atom
(assumed to be 1.4), $m_{\rm H}$ is the mass of a
hydrogen atom,
$\nh$ is the column density
of hydrogen nuclei, ${\cal D}$ is the dust-to-gas
mass ratio, and
$\langle\sigma_{\rm d}(\lambda )/m_{\rm d}\rangle$
is the cross section of dust per unit dust mass
as a function of wavelength.

Finally, we obtain the absorption line flux with
extinction, $I_{\rm line}^{\rm abs}$, as
\begin{eqnarray}
I_{\rm line}^{\rm abs}=S_\nu\Delta\nu
[1-\exp (-\tau_{\rm line})]
\exp(-\tau_{\rm dust})\, ,\label{eq:absflux}
\end{eqnarray}
where $\Delta\nu$ ($=\nu\,\Delta v/c$) is the line
width in units of frequency and $S_\nu$ is the
continuum flux (per
frequency) of the IR source behind the cloud.

We assume that all the molecular hydrogens are in
the levels $J=0,\, 1$ of the vibrational ground
state ($v=0$), since
we are interested in the clouds whose excitation
temperature is much lower than the excitation
energy. The lines are summarized in S01.
The ortho vs.\ para ratio
is assumed to be 3 : 1; i.e., $N_1=(3/4)\nmol$
and $N_0=(1/4)\nmol$, where $N_J$ is the column
density of \H2 in the
level $J$, and $\nmol $ is the total
hydrogen column density.

\begin{figure}
\begin{center}
\includegraphics[width=8cm]{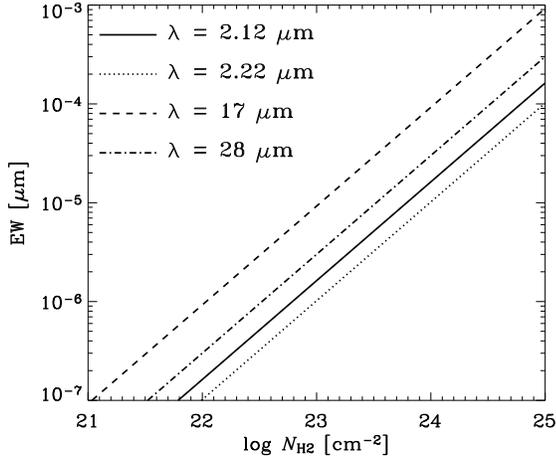}
\end{center}
\caption{Equivalent width (EW) for various
\H2 lines as a function of \H2 column density.
The solid, dotted, dashed, and
dot-dashed lines represent the line wavelengths
of $\lambda =2.12$, 2.22, 17, and 28 $\mu$m,
respectively.}
\label{fig:ew}
\end{figure}

Observationally it could be convenient to present
the equivalent width, since it is not
sensitive to the line width (or the spectral
resolution). The equivalent width $W_\lambda$
given in wavelength units
is expressed as
\begin{eqnarray}
W_\lambda & = & \frac{\lambda^2}{c}[1-\exp
(-\tau_{\rm line})]\Delta\nu\, .
\end{eqnarray}
We calculate
$W_\lambda$ as a function of $\nmol$ as
shown in
Fig.\ \ref{fig:ew}.
We should note that not only the equivalent width
but also the suppression of continuum
level by dust extinction is important for the
absorption flux (Eq.\ \ref{eq:absflux}).

\subsection{Extinction models}

The optical depth of dust, $\tau_{\rm dust}$,
is proportional to the dust-to-gas ratio,
and the hydrogen column density, if we fix
$\langle\sigma_{\rm d}(\lambda )/m_{\rm d}
\rangle$ as a function of $\lambda$ (Eq.\
\ref{eq:tau_dust}), which is taken from
Hirashita et al.\ (\cite{hirashita05}).
We adopt the Galactic extinction curve,
which is consistent with
Mathis (\cite{mathis90}), unless
otherwise stated. A different extinction curve
is examined in Sect.\ \ref{subsec:ext}.
We assume that the Galactic
dust-to-gas ratio, ${\cal D}_\odot$, is
$6\times 10^{-3}$, since we should adopt the
values in Hirashita et al.\ (\cite{hirashita05})
for consistency. We
define the normalized dust-to-gas ratio,
$\kappa$, as
\begin{eqnarray}
\kappa\equiv{\cal D}/{\cal D}_\odot\, .
\end{eqnarray}

\subsection{Background sources}\label{subsec:bg}

We focus on GRB afterglows as one of the most
luminous sources in the Universe. If some dense
clouds lie in the line of sight
of GRBs, we can have a chance to detect these
foreground clouds in absorption. For the flux
of GRB afterglows, we adopt a commonly used model
with synchrotron radiation from a relativistic
shock (Sari et al.\ \cite{sari98};
Ciardi \& Loeb \cite{ciardi00}). Here we
apply a set of values adopted by
Inoue et al.\ (\cite{inoue04}): a magnetic energy
fraction of $\epsilon_B=0.1$, an electron
energy fraction of $\epsilon_{\rm e}=0.2$, a
spherical shock energy of $E=10^{52}$ ergs, an
ambient gas number density of
$n=10~{\rm cm}^{-3}$, and a power-law index of
the electron energy distribution of $p=2.5$.
For the timescale, we put $t=1/24$ day
(1 hour), and the observational detection limits
are also calculated with 1-hour integration.

We do not include the redshift dependence of GRB
luminosity. GRBs may be brighter at high $z$ than
the local Universe
(e.g., Lloyd-Ronning et al.\ \cite{lloyd02};
Yonetoku et al.\ \cite{yonetoku04}).
In this case, detection of high-$z$ absorption
lines becomes easier.

\section{Results}\label{sec:result}

We consider a velocity width of
$\Delta v=100$ km s$^{-1}$, motivated by the
future cooled space telescopes such as
{\it SPICA} with $R\sim 3000$. Accordingly,
the frequency resolution becomes
$\Delta\nu =\nu/R$. Then, $S_\nu\Delta\nu$ is
estimated as
\begin{eqnarray}
S_\nu\Delta\nu & = & 1.0\times 10^{-18}\left(
\frac{S}{1\,{\rm mJy}}
\right)\left(\frac{\lambda}{1\,\mu{\rm m}}
\right)^{-1}\left(
\frac{R}{3000}\right)^{-1}\nonumber \\
& & ~~~~~~~~~~~{\rm W\, m}^{-2}\, .
\end{eqnarray}
We assume that the all the hydrogen atoms are in
the molecular form (i.e., $\nh =2\nmol$). This
assumption is discussed in Sect.\ \ref{subsec:gasphys}.

We note that S01 scale the extinction by $Z$
(metallicity normalized to the solar value). We have
confirmed that if we adopt $\kappa =Z$, the
difference in $\tau_{\rm dust}$ less than a factor of
2. As shown by S01, the 2.12 $\mu$m absorption
line is the strongest, and the dust obscuration does
not affect the line detection if $\kappa\la 0.01$.
Therefore, we concentrate on
the 2.12 $\mu$m line with $\kappa =0.01$ in this
section.

As stated by S01, it is quite difficult to detect
absorption lines where optical depth is less than
0.01, since an extremely high signal-to-noise
ratio would be required to detect an absorption
with $\tau_{\rm line}<0.01$. The \H2 column
density which satisfies $\tau_{\rm line}=0.01$
is listed in S01 for various lines. For the
2.12 $\mu$m line, $\nmol$ should be larger than
$4\times 10^{23}~{\rm cm}^{-2}$
to satisfy $\tau_{\rm line}>0.01$.
Therefore, if we use the NIR \H2 lines,
we should target the absorption systems with
$\log\nmol\ga 23.5$. However, we should
note that this condition is dependent on the
line width. The sound speed of \H2-rich cold
clouds is generally much less than
100 km s$^{-1}$. Therefore, if we had a
facility sensitive enough with a resolution
large enough,
we could observe an absorption system whose
\H2 column density is smaller than
$10^{23.5}~{\rm cm}^{-2}$.

\begin{figure}
\begin{center}
\includegraphics[width=8cm]{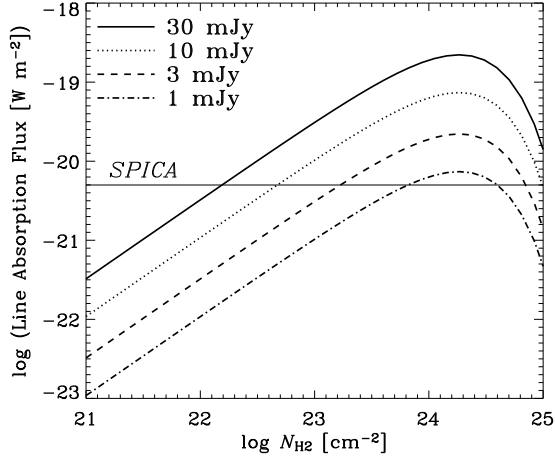}
\end{center}
\caption{2.12 $\mu$m line absorption flux as a
function of the column density of hydrogen
molecules. Various continuum fluxes of the
background sources are examined. The solid,
dotted, dashed, and dot-dashed curves
correspond to the continuum fluxes of
30, 10, 3, and 1 mJy, respectively.
The horizontal thin solid line shows the
expected detection limit of future cooled
space telescopes such as {\it SPICA}
(5 $\sigma$ for an integration of
$\sim 1$ hour).}
\label{fig:varflux}
\end{figure}

In Fig.\ \ref{fig:varflux}, we show the
2.12 $\mu$m absorption fluxes as a function of
$\nmol$ for various background fluxes. The typical
detection limit expected for future IR cooled
telescopes such as {\it SPICA} is also shown
($5\times 10^{-21}$ W m$^{-2}$ for 5 $\sigma$
detection with an integration of $\sim 1$ hour;
Ueno et al.\ \cite{ueno00}; S01) as a
representative observability of future facilities.
With future space missions, high column density
systems with
$\log \nmol \sim 23.5$--25 can be
investigated if the background source
luminosity is $\ga 10$ mJy.

\begin{figure}
\begin{center}
\includegraphics[width=8cm]{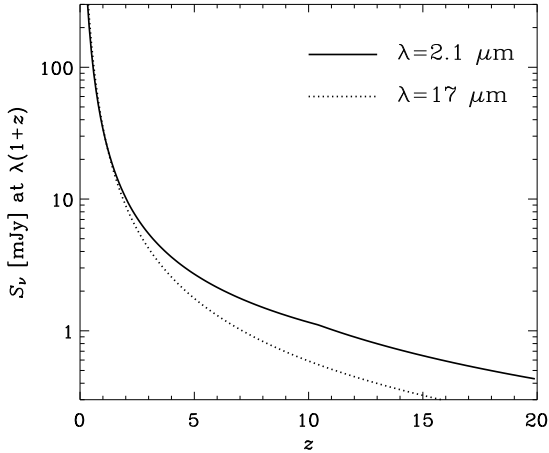}
\end{center}
\caption{Observed flux of a gamma-ray burst
afterglow as a function of redshift. The
solid and dotted lines
correspond to the restframe wavelengths of
2.12 and 17 $\mu$m, i.e., observational
wavelengths of $2.12(1+z)$ and
$17(1+z)~\mu$m, respectively.}
\label{fig:grbflux}
\end{figure}

In order to show the flux level of GRB afterglows,
we present Fig.\ \ref{fig:grbflux}. The solid and
dotted lines show the flux ($S_\nu$)
at the restframe wavelengths of
$\lambda =2.1~\mu$m and 17 $\mu$m (i.e., the
observational wavelengths of
$2.1(1+z_{\rm GRB})~\mu$m and
$17(1+z_{\rm GRB})~\mu$m), respectively, where
$z_{\rm GRB}$ is the redshift of the GRB afterglow.
We observe that the afterglows are more luminous
at $2.1(1+z_{\rm GRB})~\mu$m than at
$17~(1+z_{\rm GRB})~\mu$m. In reality, we should
observe at $2.1(1+z_{\rm abs})~\mu$m or
$17(1+z_{\rm abs})~\mu$m, where $z_{\rm abs}$
is the redshift of a target absorption line system,
but if $z_{\rm abs}$ has the same order of
magnitude as $z_{\rm GRB}$, the absorption flux
does not change very much
(the dependence on other parameters assumed in
Section \ref{subsec:bg} is larger).
Considering that the 2.1 $\mu$m absorption line is
stronger than the other lines, observations at
$2.1(1+z_{\rm abs})~\mu$m is
favorable to trace high column density
($\log \nmol \sim 23.5$--25) systems (see S01
for the other lines).

\begin{figure}
\begin{center}
\includegraphics[width=8cm]{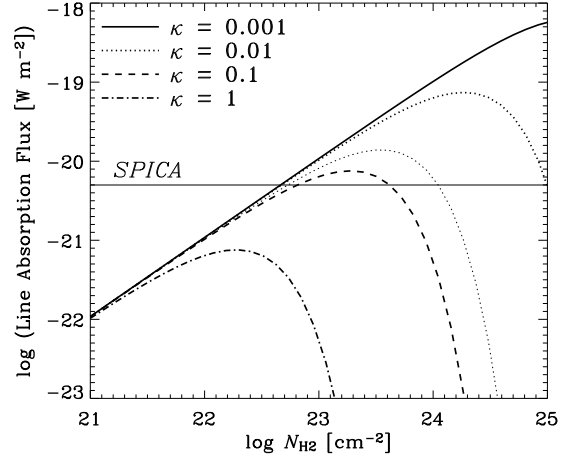}
\end{center}
\caption{Same as Fig.\ \ref{fig:varflux}, but for
various dust-to-gas ratio
($\kappa$ is the dust-to-gas ratio normalized
to the Galactic value). The continuum fluxes
of the background sources is assumed to be
10 mJy. The
solid, dotted, dashed, and dot-dashed lines
correspond to $\kappa =0.01$, 0.1, 1, and
0.001, respectively. The lower dotted line
represents the result with the H05b extinction
curve, while the other lines adopt the
Galactic extinction curve.}
\label{fig:vardg}
\end{figure}

The dependence on the dust-to-gas ratio is also
shown for the background flux of 10 mJy
(Fig.\ \ref{fig:vardg}). The solid, dotted, dashed,
and dot-dashed lines represent the absorption line
fluxes with $\kappa =0.0001$, 0.001, 0.01, and 1,
respectively. We see that if $\kappa >0.1$, we
can detect the NIR \H2 line if
$\log \nmol \ga 23$.

\section{Discussions and observational
strategies}\label{sec:strategy}

\subsection{High column density systems
in cosmological scenarios}
\label{subsec:gasphys}

\subsubsection{Protogalactic disks}

Collapsed gas in a dark matter halo may
form a disk because of the initial angular momentum.
If we observe
a GRB in a nearly edge-on disk, we could identify the
disk as a high column density system.
If a galactic disk has a typical circular velocity of
100 km s$^{-1}$, the typical mass ($M_{\rm disk}$)
and radius ($r_{\rm disk}$) of the disk should be
$M_{\rm disk}\sim 2\times 10^9~M_\odot$ and
$r_{\rm disk}\sim 6$ kpc if it forms at $z\sim 3$
(Hirashita et al.\ \cite{hirashita03}). With these
mass and size, the typical hydrogen column density
along the galactic disk is
$\sim n_{\rm H}r_{\rm disk}$. Using the relation
$M_{\rm disk}\sim\pi r_{\rm disk}^2H\mu
m_{\rm H} n_{\rm H}$, where $H$ is the disk thickness,
we obtain
\begin{eqnarray}
n_{\rm H}r_{\rm disk} & \sim &
1\times 10^{23}\left(
\frac{M_{\rm disk}}{2\times 10^9~M_\odot}\right)
\left(\frac{r_{\rm disk}}{6~{\rm kpc}}\right)^{-1}
\nonumber \\
& \times & \left(\frac{H}{100~{\rm pc}}\right)^{-1}
~{\rm cm^{-2}}\, .
\end{eqnarray}
Therefore, a typical galactic disk could be observed
as a high column density system if a GRB lies within
or behind the disk and if a significant fraction of
the gas is in molecular form.

As stated in Sect.\ \ref{sec:result}, it is required
that the dust-to-gas ratio is $\la 10^{-2}$ times
the Galactic value (i.e.,
${\cal D}\la 6\times 10^{-5}$). With the typical disk mass
mentioned above ($2\times 10^9~M_\odot$), the corresponding
dust mass is $M_{\rm dust}\la 10^5~M_\odot$. We estimate
the timescale of the dust enrichment up to
this level. Galactic disks observed as DLAs
can be considered as objects with a star formation rate
of $\sim 0.1-1~M_\odot~{\rm yr}^{-1}$
(Hirashita \& Ferrara \cite{hf05}, hereafter HF05).
Using the Salpeter
initial mass function, we obtain the supernova rate of
$\sim 10^{-3}-10^{-2}$ yr$^{-1}$. Assuming that
around 0.5 $M_\odot$ of dust is formed in a SN
(Todini \& Ferrara \cite{todini01};
Nozawa et al.\ \cite{nozawa03}), we obtain a typical
dust formation rate of
$\sim 5\times 10^{-4}-5\times 10^{-3}~M_\odot$ yr$^{-1}$.
Thus, in
order to accumulate the dust mass estimated above, a
timescale
of $2\times 10^7-2\times 10^8$ yr is required, which
is much
larger than the lifetime of the GRB progenitors
(Heger et al.\ \cite{heger03}).

It is observationally known that some DLAs have
dust-to-gas ratio of $\kappa\sim 10^{-2}$
(Pettini et al.\ \cite{pettini97};
Ledoux et al.\ \cite{ledoux03}; HF05).
This observational evidence strongly suggests that
galactic disks with $\kappa\sim 10^{-2}$ exist
at high redshift.

\subsubsection{Supergiant molecular complexes}
\label{subsubsec:giant}

In some starburst galaxies such as the Antennae
(Wilson et al.\ \cite{wilson03}) and Arp 229
(Casoli et al.\ \cite{casoli99}),
supergiant molecular complexes are observed.
In those galaxies, molecular gas of
$\sim 10^9~M_\odot$ is concentrated in a 1-kpc$^2$
area. This indicates an \H2 column density as large
as $\sim 10^{23}~{\rm cm}^{-2}$ (and a hydrogen
number density of $n_{\rm H}\sim 30$ cm$^{-3}$).
Arp 220 seems to have a comparable amount of gas in
more concentrated region
(Sakamoto et al.\ \cite{sakamoto99}). Thus, the
\H2 column density in a starbursting region can exceed
several times $10^{23}~{\rm cm}^{-2}$.
This column density meets the requirement to detect
\H2 in absorption. 

\begin{figure}
\begin{center}
\includegraphics[width=8cm]{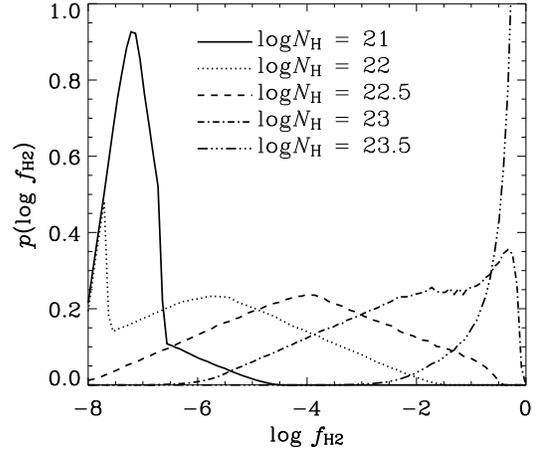}
\end{center}
\caption{Probability distribution function of the molecular
fraction ($\log\fH2$) for various hydrogen column
densities. The solid, dotted, dashed, dot-dashed, and
dot-dot-dot-dashed lines correspond to
$\log \nh~({\rm cm}^{-2})=21$, 22,
22.5, 23, and 23.5, respectively. The dust-to-gas ratio
normalized to the Galactic value is assumed to be
$\kappa =0.01$.}
\label{fig:pfH2}
\end{figure}

HF05 have recently developed a method to calculate a
probable range of $\fH2$ under a given range of
physical quantities concerning ISM. Their assumption
is that $\fH2$ is determined by the equilibrium
between formation on dust grains
and destruction by dissociating photons.
The probability density function of $\log\fH2$,
$p(\log\fH2 )$ is calculated by using the
method described in Appendix B of HF05. 
As a first guess, we start from the ranges of
quantities derived for \H2-detected
DLAs by HF05:
$1.5<\log n~[{\rm cm}^{-3}]<2.5$,
$1.5<\log T~[{\rm K}]<3$, and
$0.5<\log\chi <1.5$, where $n$ is the
number density of gas particles, $T$ is
the gas temperature, and $\chi$ is the
UV interstellar radiation field
intensity normalized to the Galactic value
taken from Habing (\cite{habing68}).
We adopt $\kappa =0.01$ for
the dust-to-gas ratio
(Sect.\ \ref{sec:result}).
In Fig.\ \ref{fig:pfH2}, we
show $p(\log\fH2 )$ for various $\nh$.
{}From Fig.\ \ref{fig:pfH2}, we observe that
the gas becomes almost fully molecular for
$\log\nh\ga 23$ even if the dust-to-gas ratio
is only 1 percent of
the Galactic value ($\kappa\sim 0.01$). This
is consistent with the conclusion
by Schaye (\cite{schaye01}) that the large part
of hydrogen is in molecular form at
$\log\nh\ga 22$.
We stress that at $\kappa\sim 0.01$
\H2 formation is
accelerated and possibly leads to a burst of
star formation
(Hirashita \& Ferrara \cite{hf02}).
Therefore, our proposed observation has an
importance that we trace the reservoir of gas
clouds which might cause a starburst.
Indeed Iliev et al.\ (\cite{iliev05}) investigate
the fate of those clouds proposed by HF05
and conclude that
such clouds
can finally collapse even in the presence of
UV radiation.

\subsection{Extinction curves}
\label{subsec:ext}

Some observations show that the extinction curves of
high-$z$ galaxies are different from local ones
(Maiolino et al.\ \cite{maiolino04}).
Therefore, it is interesting to examine the
extinction curve derived for high-$z$ galaxies.
We adopt the theoretical curve of
Hirashita et al.\ (\cite{hirashita05}).
We adopt Model b (based on the dust production in
an unmixed SNe II; Nozawa et al.\ \cite{nozawa03})
of Hirashita et al.\ (\cite{hirashita05}), which
nearly reproduce the observational high-$z$
extinction curve of
Maiolino et al.\ (\cite{maiolino04}).
This model is labeled
as H05b (see Takeuchi et al.\ \cite{takeuchi05}
for the NIR part of the extinction curve).

In Fig.\ \ref{fig:vardg}, we show the absorption
flux of the 2.12~$\mu$m line
for the H05b extinction curve by the lower
(thinner) dotted line, where the dust-to-gas
ratio is assumed to be
$\kappa =0.01$ (0.01 times the Galactic value).
The upper (thicker) dotted line corresponds to
the Galactic extinction curve with the same
dust-to-gas ratio.
The difference in the extinction curve does not
affect the absorption line flux for
$\log \nmol \la 23$, since the
optical depth of dust is small, but
the difference in the extinction becomes
significant for $\log \nmol \ga 23.5$.
In H05b the system becomes more opaque for
$\log\nmol\ga 24$ than in the Galactic
extinction case.
This sensitive dependence on the extinction
could also be used to
constrain the IR extinction curves at
high $z$ (see also
Vergani et al.\ \cite{vergani04}).

\subsection{Number Estimate}
\label{subsec:future}

There is no observational constraint on the
number of high column density systems with
$\log\nh\ga 22$. One simple way to
estimate the number is to extrapolate the
distribution of $\nh$ sampled up to
$\log\nh\sim 22$, although
Vladilo \& P\'{e}roux (\cite{vladilo05}) argue
that the distribution should not be
extrapolated in the range of column densities
not sampled by the observation.
After correcting the
distribution for dust extinction,
Vladilo \& P\'{e}roux (\cite{vladilo05})
indicate that
$\log f(\nh\sim 10^{22}~{\rm cm^{-2}})
\sim -23.5$,
where $f(\nh )\,\rd\nh\,\rd X$
represents the number of absorbers along
a random line of sight with hydrogen column
densities between $\nh$ and
$\nh +\rd\nh $ and redshift paths
between $X$ and $X+\rd X$
(e.g., P\'{e}roux et al.\ \cite{peroux03}).
By assuming $f(\nh )\propto\nh^{-\beta}$
with $\beta =1.5$
(Vladilo \& P\'{e}roux \cite{vladilo05}),
we obtain $\log f(\nh )\sim -25.8$
at $\log\nh\sim 23.5$.
At $z=2$, $X\simeq 4.4$ and if we take the
interval of $\nh $ as
$\sim 10^{23}~{\rm cm}^{-2}$, we obtain
the number of absorbers in a random line of
sight as $\sim 10^{-2.8}$. Therefore, around
600 lines of sight are necessary to detect
a high column density object. If a large
part of those sources are favorable for
\H2 formation as suggested in
Sect.\ \ref{subsec:gasphys},
we can detect \H2 lines every $\sim 600$
GRBs.

For example, the sample of
Schmidt (\cite{schmidt01}) consists of 1391 GRBs
in a period of 5.9 yr. Their analysis indicates
that about 500 GRBs are around
$z\sim 2$. Therefore,
it is possible that about 100 GRBs are
identified around $z\sim 2$ per year. Therefore,
we could find 0.2 GRBs
associated with high column density systems
per year. If we consider
an improvement of the sensitivity in the future,
we expect a larger number.

\section{Conclusions}\label{sec:conclusion}

In this paper, we have investigated the
possibility to detect IR \H2 absorption lines from
dust-poor high column density ($\log \nh \ga 23.5$)
clouds at high-$z$. \H2 is a unique tracer of high
column density clouds, because hydrogen is
considered to be in molecular form even at
extremely low dust-to-gas ratio ($\kappa\sim 0.01$,
where $\kappa$ is the dust-to-gas ratio normalized
to the Galactic value)
(Sect.\ \ref{subsec:gasphys}). We have shown that if
$\kappa < 0.1$, we can observe \H2 absorption lines
of gas whose column density is $\log\nmol \ga 23.5$.
In particular,
if $\kappa\la 0.01$, clouds with
$\log \nmol \sim 23.5$--25 are accessible
(Sect.\ \ref{sec:result}; Fig.\ \ref{fig:vardg}).
Since near future
missions such as {\it SPICA} will access GRB
afterglows up to $z\sim 2$ in the IR regime, the
search for metal-poor (or dust-poor) galaxies at
$z\la 2$ is an important scientific target of near
future IR telescopes (Sect.\ \ref{subsec:future}).
We should note that some bright GRBs at high-$z$
can be utilized to trace higher-$z$ primeval galaxies.

\begin{acknowledgements}

We are grateful to A. Ferrara, C. P\'{e}roux, P. Petitjean,
and T. Nakamura for stimulating discussions.
We acknowledge several comments of the anonymous referee.
H. H. and T.\ T.\ T.\ have been supported
by the Research Fellowship of the Japan Society for the Promotion
of Science for Young Scientists.

\end{acknowledgements}


\end{document}